\documentclass[prb,superscriptaddress,twocolumn,showpacs]{revtex4-1}
\usepackage{amsmath}
\usepackage{amsfonts}
\usepackage{amssymb}
\usepackage{graphicx,graphics,color}

\newcommand{\beq}{\begin{equation}}
\newcommand{\eeq}{\end{equation}}
{
 \definecolor{BLACK}{gray}{0}
 \definecolor{WHITE}{gray}{1}
 \definecolor{RED}{rgb}{1,0,0}
 \definecolor{GREEN}{rgb}{0,1,0}
 \definecolor{BLUE}{rgb}{0,0,1}
 \definecolor{CYAN}{cmyk}{1,0,0,0}
 \definecolor{MAGENTA}{cmyk}{0,1,0,0}
 \definecolor{YELLOW}{cmyk}{0,0,1,0}
}

\begin{document}

\title{Superconductivity at low density near a ferroelectric quantum
critical point: doped SrTiO$_{3}$ }
\author{Peter W\"{o}lfle}
\author{Alexander V. Balatsky }
\date{\today }

\begin{abstract}
Recent experiments on electron- or hole-doped SrTiO$_{3}$ have revealed a
hitherto unknown form of superconductivity, where the Fermi energy of the
paired electrons is much lower than the energies of the bosonic excitations
thought to be responsible for the attractive interaction. We show that this
situation requires a fresh look at the problem calling for (i) a systematic
modeling of the dynamical screening of the Coulomb interaction by ionic and
electronic charges, (ii) a transverse optical phonon mediated pair
interaction and (iii) a determination of the energy range over which the
pairing takes place. We argue that the latter is essentially given by the
limiting energy beyond which quasiparticles cease to be well defined. The
model allows to find the transition temperature as a function of both, the
doping concentration and the dielectric properties of the host system, in
good agreement with experimental data. The additional interaction mediated
by the transverse optical soft phonon is shown to be essential in explaining
the observed anomalous isotope effect. The model allows to capture the
effect of the incipient (or real) ferroelectric phase in pure, or oxygen
isotope substituted SrTiO$_{3}$ .
\end{abstract}

\maketitle

\affiliation{Institute for Theory of
Condensed Matter, Karlsruhe Institute of Technology, 76049 Karlsruhe, Germany}
\affiliation{Institute for Nanotechnology, Karlsruhe Institute of
Technology, 76021 Karlsruhe, Germany}

\affiliation{Nordita, Stockholm, SE 10691, Sweden}

\affiliation{Institute for Condensed Matter Theory and Institute for Nanotechnology,
Karlsruhe Institute of Technology, 76021 Karlsruhe, Germany}

\affiliation{Nordita, Stockholm, SE 10691, Sweden}

\section{Introduction}

The question of the nature of the superonducting state in SrTiO$_{3}$ (STO)
recently has been sharply brought into focus. There are at least two main
issues driving the current debate about superconductivity in STO. One is the
observation of the superonducting state at extremely low dopings, well below
those of any other known superconductors. Recent observations of
superconductivity in STO at such low doping levels \cite{Behnia13,Behnia14}
that the Fermi energy is much less than the relevant phonon energies are
calling for a fresh look at the theory of superconductivity, even in a
so-called weakly correlated material. Another issue that is clearly present
is the interplay between superconductivity and the ferroelectric quantum
critical point in STO. STO is a "quantum paraelectric" close to the quantum
phase transition into the ferroelectric state \cite{Mueller79}, and may be
tuned into the ordered phase, e.g. by isotope substitution of $^{16}$O\ by $%
^{18}$O\ \cite{Rowley14} or by substituting Ca for Sr \cite{Bednorz84}. It
becomes a metal by substituting, e.g. Sr with La, or Ti with Nb, or by
removing O. These two questions: electron-electron interactions and the
ferroelectric quantum critical point (QCP) are often intermingled in current
discussions, yet they represent very different physics that might be
connected at the end but does not have to be. Hence we start by discussing
pairing due to dynamically screened electron-electron interaction, which is
only weakly dependent on the proximate ferroelectric quantum criticality. It
is, however, strongly modified by the large static dielectric constant. We
thus will first focus on the role of screening and pairing at low doping
levels in STO. Then we advocate the importance of an additional interaction
sensitive to the ferroelectric fluctuations and giving rise to a strong
isotope effect.

Superconductivity in STO metal at low temperatures has been observed long
ago \cite{Schooley64}, the transition temperature as a function of doping
peaking at a maximum of $T_{c}\approx 0.4$K at a carrier density of $%
n\approx 10^{20}$cm$^{-3}$ \cite{Koonce67,Binnig80,Binnig81}. According to
the conventional view of superconductivity the main question is to identify
the "glue" binding conduction electrons into a Cooper pair. An attractive
interaction component may be obtained from suitable boson exchange
processes: phonons, spin fluctuations, plasmons, etc. The remaining Coulomb
repulsion is often neglected, on the grounds that it is substantially
reduced in magnitude if downfolded into the low energy regime where the
exchange bosons live. While this line of argumentation works reasonably well
for superconductors with well separated energy scales, such that the Fermi
energy $\epsilon _{F}>>\omega _{boson}$ , a typical boson energy, it breaks
down in case that the energy scales are no longer separated. It is then
necessary to treat dynamical screening by ionic charges and by conduction
electrons on the same footing. Early theories of superconductivity of STO ,
e.g. Koonce et al. \cite{Koonce67} and Appel \cite{Appel69} , and even
recent theories \cite{Gorkov16}\ did not address this problem properly and
are therefore not suitable in the low doping domain $n<10^{20}$cm$^{-3}$. A
later theory by Takada \cite{Takada80}\ focusses on the interplay of ionic
and electronic screening, starting from a pair interaction given by the
screened\ Coulomb interaction (see below) and using the pairing theory in
terms of the dynamical dielectric constant \cite{Kirshnits73}. Takada uses
additional approximations such as a plasmon pole approximation and a
frequency cutoff of the order of the Fermi energy, without justification
(for a discussion of the proper cut-off see below and Appendix A)), so that
his results are questionable. A similar starting point has been chosen
recently by Ruhman and Lee \cite{Ruhman16}, who again use additional
approximations (treatment of the Coulomb interaction at high frequency, and
energy cut-off at the Fermi energy) without sufficient justification. The
latter authors identify at least two problems with this more conventional
approach: (i) by employing the Fermi energy as an energy cutoff the relevant
interaction regime in energy space is found to shrink much too fast for
decreasing density, leading to tiny transition temperatures; (ii) even at
higher densities the T$_{c}$ values come out too low, so that the authors
propose that the extremely high dielectric constant found in the undoped
system is substantially reduced by doping, which helps to increase the
screened Coulomb interaction and thus, T$_{c}$. Experimentally there is no
indication that the very large dielectric constant is substantially reduced
by doping at the low levels in question here. Both problems are resolved
below by applying the proper cutoff and by taking into account an additional
attractive pair interaction mediated by transverse optical (TO) phonons.

A careful treatment of the frequency cutoff $\omega _{c}$\ is all-important
in this problem, since here the cutoff frequency enters the result for the
transition temperature in an essential way. We argue below that the
processes limiting the pairing to a low energy regime in the case that the
Fermi energy is less than the typical phonon energy are given by the
self-energy entering the anomalous Green's function in the gap equation.
While in our present solution of the gap equation we do not fully include
the self-energy, we approximate its effect by introducing an energy cutoff
at the energy beyond which quasiparticles are no longer well defined. In
Appendix A we estimate this cutoff in three separate density regimes, with
different dependencies on density in each regime. At the lowest densities
electron-phonon interaction causes a vanishing of the cutoff with falling
density $n$, $\omega _{c}\propto n^{1/3}$, while at the highest densities
Fermi liquid theory provides $\omega _{c}\propto n^{2/3}$. The surprising
result of the study of the quasiparticle relaxation rate (see Appendix A) is
that at intermediate densities, where $\epsilon _{F}\ll \omega _{LO}$, the
longitudinal optical phonon energy, the cutoff is found to have a reversed
trend, $\omega _{c}\propto n^{-1/6}$, leading to a nonmonotonic dependence
of $\omega _{c}(n)$.

In the conventional theory of superconductivity the small parameter $\omega
_{phon}/\epsilon _{F}$ allows to show that higher order correction terms
(vertex corrections, crossed diagrams, etc) are small and may be safely
neglected (Migdal's theorem). In the case of STO the presence of different
small parameters again allows to neglect higher order corrections: the ratio
of the transverse optical (TO, soft mode) and the longitudinal optical (LO)
phonon frequencies squared, $(\omega _{TO}/\omega _{LO})^{2}$, which is of
order $10^{-3}$ at wavevectors $q=\kappa $, the screening wave vector, and
the weak coupling of electrons to TO phonons. The dimensionless Coulomb
interaction, $N_{F}V_{C}$ is of order $10^{-2}$ in the relevant energy and
wavevector domain, rather than of order unity as in conventional metals
(here $N_{F}$ is the density of states at the Fermi level). As a consequence
vertex corrections are small, as estimated in Appendix C. Moreover, the
nominally small density of carriers might suggest that the electron system
is in the low density, strong interaction regime. The opposite is the case:
the effective Bohr radius $a_{B}^{\ast }$ is a factor of $\varepsilon
_{0}(m/m_{1})\approx 10^{4}$ larger than in usual metals, such that the
parameter $r_{s}^{\ast }=(3n/4\pi )^{-1/3}/a_{B}^{\ast }\approx 0.01$, which
puts the system into the effectively high density, weakly coupled regime.\
The additional interaction mediated by TO-phonon exchange is on one hand
small due to the in general small coupling of transverse phonons to
electronic quasiparticles, but is on the other hand boosted by the unusual
screening properties arising with the soft TO-phonon mode, leading to large
dynamical effective charges. Nonetheless, the vertex corrections induced by
the TO-phonon mediated interaction are again small, as estimated in Appendix
C.

The role of ferroelectric fluctuations is prominent in the observed gigantic
isotope effect, occurring when the usual oxygen isotope O$^{16}$ is replaced
by the heavier O$^{18}$. In conventional BCS-theory this substitution should
reduce the transition temperature $T_{c}$ by a few percent. What is observed
is, however, a drastic increase of $T_{c}$ by as much as $50\%$ for an O$%
^{18}$ concentration of $x=0.35$ \cite{vdMarel16}. As mentioned above, the
isotope substitution moves STO towards the ferroelectric phase, as signaled
by the divergence of the paraelectric susceptibility and the concomitant
vanishing of the TO-phonon soft mode frequency $\omega _{TO}(q=0)$. The
latter is expected to boost the contribution of the TO-phonon mediated pair
interaction, as is indeed found (see below). The fundamental effect of the
critical ferroelectric fluctuations on the superconductivity in STO has been
proposed early on by \cite{Edge15,Kedem16}. These authors employed a model
of quantum criticality featuring a soft mode, which may be identified as the
TO phonon mode. In a phenomenological model they estimated the effect of a
pairing interaction mediated by exchange of the soft mode on the transition
temperature and found a significant effect, as found later in experiment
\cite{vdMarel16}.

What is the interaction between two conduction electrons in a metal, really?
On the one hand it is given by the dynamically screened Coulomb interaction $%
V_{C}$, including any process contributing to the screening, phonons,
plasmons, or any other coherent or incoherent type of excitation. This fully
screened interaction between two charges in the solid has to be repulsive in
the static limit, for reasons of stability. As will be shown below, in
Matsubara frequency space the interaction is always positive, $V_{C}(\mathbf{%
q},i\omega _{n})>0$, where $\mathbf{q,}\omega _{n}$ is the momentum and
energy transferred in the interaction process. The second type of
interaction is of exchange character, and is thought to provide the pairing
"glue" for many unconventional superconductors such as the cuprates, the
pnictides and the heavy fermion compounds, often by exchange of spin
fluctuations. We will show below that in the present case of a soft
transverse phonon mode, exchange of these excitations may make a substantial
contribution to the pairing interaction.

\section{Pair interaction.\emph{\ }}

The pair interaction thus consists of two parts
\begin{equation}
V_{pair}(\mathbf{q},i\omega _{n})=V_{C}(\mathbf{q},i\omega _{n})+V_{soft}(%
\mathbf{q},i\omega _{n})
\end{equation}%
Here the dynamically screened Coulomb interaction is given by
\begin{equation}
V_{C}(\mathbf{q},i\omega _{n})=\frac{4\pi e^{\ast 2}}{\varepsilon (\mathbf{q}%
,i\omega _{n})q^{2}},
\end{equation}%
with the dielectric function capturing the screening by electronic and ionic
charges

\begin{equation}
\varepsilon (\mathbf{q},i\omega _{n})=[1+\frac{4\pi e^{\ast 2}}{q^{2}}\chi
_{el}(\mathbf{q},i\omega _{n})]\frac{(i\omega _{n})^{2}-\omega _{LO}^{2}(%
\mathbf{q},i\omega _{n})}{(i\omega _{n})^{2}-\omega _{TO}^{2}(\mathbf{q})}.
\end{equation}%
The dielectric function vanishes at $i\omega _{n}=\omega _{LO}(\mathbf{q}%
,i\omega _{n})$ , defining the longitudinal optical phonon frequency, and
diverges at the transverse optical phonon frequency, $i\omega _{n}=\omega
_{TO}(\mathbf{q})$. The electronic screening effect is embodied in the
irreducible electronic charge susceptibility $\chi _{el}$. In the undoped
limit $\chi _{el}\rightarrow 0$.\ The effect of higher lying excitations is
lumped into the optical dielectric constant $\varepsilon _{\infty }$,
renormalizing the electron charge as $e^{\ast }=e/\sqrt{\varepsilon _{\infty
}}$ . We employ a single mode model, recognizing the fact that the TO-phonon
mode becoming soft in the nearly ferroelectric SrTiO$_{3}$, and its
longitudinal partner are the dominant modes . If necessary the model may be
generalized to include further phonon modes. For simplicity we consider a
fully isotropic model, approximating the ellipsoidal Fermi surfaces \cite%
{vdMarel11} by spheres. The phonon properties are actually very anisotropic
(see Appendix B), but we will approximate the phonon dispersion by an
angular average as well.

We may then parametrize the phonon frequencies as (for details see Appendix
B)

\begin{eqnarray}
\omega _{TO}^{2}(\mathbf{q}) &=&\omega _{D}^{2}(\tau +\lambda q^{2}),  \notag
\\
\omega _{LO}^{2}(\mathbf{q},i\omega _{n}) &=&\omega _{TO}^{2}(\mathbf{q})+%
\frac{4\pi \omega _{D}^{2}q^{2}}{q^{2}+4\pi e^{\ast 2}\chi _{el}(\mathbf{q}%
,i\omega _{n})}
\end{eqnarray}%
in the interval $0<q<q_{c}$. The parameter $\tau $ is closely related to the
dielectric constants of the host, $\varepsilon _{0}=\varepsilon _{\infty
}(1+4\pi /\tau )$. The static dielectric constant at $q=0$, is very large, $%
\varepsilon _{0}\approx 2\times 10^{4}$ at Kelvin temperatures \cite%
{Weaver59} and the optical dielectric constant is $\varepsilon _{\infty
}\approx 5.2$ \cite{Spitzer62,Kamaras95}. Substituting these values, we find
$\tau \approx 3.27\times 10^{-3}$. We note that\ $\tau $ acts as a control
parameter of the quantum phase transition to the ferroelectric phase,
tending to zero at the transition, $\tau (x)=\tau (1-x/x_{c})\rightarrow 0$
at $x=x_{c}$, where $x$ is the concentration of O$^{18}$ in the case of
isotope substitution. The analysis of Raman scattering data \ \cite{Vogt95}
and of inelastic neutron scattering data \cite{Yamada69}, on the zone-center
soft-mode transverse optical phonon allows to extract the following values
of the above parameters: $\omega _{TO}(\mathbf{q\rightarrow 0})\approx 1.9$%
meV ($22$K in Kelvin units), which using $\tau $ gives a characteristic
phonon energy $\omega _{D}\approx 33$meV ($380$K). The parameter $\lambda $
averaged over the direction of $\mathbf{q}$ is found as $\lambda \approx 4.38%
\mathring{A}^{2}$ (see Appendix B).

With five atoms in the unit cell, SrTiO$_{3}$ has $15$ phonon modes in
total. The three acoustic modes are less relevant for the pairing and will
be discarded. Of the remaining four triplets of optical phonon modes one is
not infrared active, i.e. does not couple well to electrons, and may be
omitted. The most relevant triplet consists of the two degenerate TO soft
modes, which are surprisingly correlated with the highest lying longitudinal
mode (usually called LO4)\cite{Vanderbilt94}. This is the set of phonon
excitations which is well described by a phenomenological
Ginzburg-Landau-Wilson model of the dynamic electric polarization of the
system as shown in Appendix B. We neglect the remaining two triplets of
modes. The experimental observation of a relatively strong e-ph coupling of
the highest LO phonon mode \cite{Mannhardt15,Swartz18} seen in tunneling
experiments is compatible with our model, considering that real rather than
virtual phonon excitations are involved.

We approximate $\chi _{el}$ by its non-interacting limit for an isotropic
system, given by

\begin{equation}
\chi _{el}^{(0)}(\mathbf{q},i\omega _{n})=2\int \frac{d^{d}k}{(2\pi )^{d}}%
\frac{n(\epsilon _{\mathbf{k+q}})-n(\epsilon _{\mathbf{k}})}{i\omega
_{n}-\epsilon _{\mathbf{k+q}}+\epsilon _{\mathbf{k}}},
\end{equation}%
where $\epsilon _{\mathbf{k}}=\frac{k^{2}}{2m_{1}}$\ , with $m_{1}\approx
1.8m$ a Fermi surface average of the effective mass of the lowest band
(assuming $\epsilon _{\mathbf{k}}\ll W$, bandwidth) , and $n(\epsilon _{%
\mathbf{k}})=[1+e^{(\epsilon _{\mathbf{k}}-\mu )/T}]^{-1}$ is the Fermi
function (we use units for which Planck's constant $\hbar $ and Boltzmann's
constant $k_{B}$ are equal to unity). The chemical potential $\mu $ at
doping density $n$ will be determined in the low temperature limit as $\mu
=\epsilon _{F}=k_{F}^{2}/2m_{1}$, where $k_{F}=(3\pi ^{2}n)^{1/3}$,
considering that for the low doping levels in question here only states near
the bottom of the lowest band are of interest. It will be seen later that
the temparatures of interest satisfy $k_{B}T\ll \mu $. In case of several
occupied electronic bands and effective mass renormalizations the expression
for $\chi _{el}$ may be generalized correspondingly. The above assumes a
rigid band model, i.e. the effect of doping is simply to populate the
unoccupied conduction bands of SrTiO$_{3}$ (Ti derived $d-$bands).

The TO-phonon mediated interaction takes the form
\begin{equation}
V_{soft}(\mathbf{q},i\omega _{n})=-2|M_{\mathbf{q}}^{TO}(\mathbf{k})|^{2}%
\frac{2\omega _{TO}(q)}{\omega _{n}^{2}+\omega _{TO}^{2}(q)}
\end{equation}%
and adds an attractive contribution to $V_{pair}$ (the factor of $2$
accounts for the two nearly degenerate soft TO modes). Here $M_{\mathbf{q}%
}^{TO}(\mathbf{k)}$ is the electron-phonon coupling function, the transition
amplitude for scattering of a conduction electron from momentum state $%
\mathbf{k}$ to state $\mathbf{k+q}$\ by emission of a TO phonon (we consider
only the one mode going soft at the QCP; we also neglect umklapp processes).
In deformation potential approximation it is given by

\begin{equation}
M_{\mathbf{q}}^{TO}(\mathbf{k})=-i\sqrt{\frac{\hbar N/V}{2m_{ion}\omega _{TO}%
}}\sum_{\alpha }\frac{(\mathbf{q\cdot e}_{TO,\alpha })U_{\mathbf{q,}\alpha }%
}{1+(\kappa /q_{typ})^{2}}B_{\mathbf{q}}(\mathbf{k})  \label{MTO}
\end{equation}%
where $m_{ion}$ is an effective ion mass,\ and $U_{\mathbf{q,}\alpha }$ are
the Fourier components of the Coulomb potential of ion $\alpha $\ inside a
unit cell, $U_{\mathbf{q},\alpha }=\int_{uc}d^{3}re^{i\mathbf{qr}}U_{\alpha
}(\mathbf{r)}$. The soft TO phonon mode is essentially a vibration of oxygen
against the titanium ion. We therefore single out these two ions and assume
the displacement vectors to be equal and opposite, $\mathbf{e}_{TO,Ti}=-%
\mathbf{e}_{TO,O}=\mathbf{e}_{TO}$. This leads to a total deformation
potential of $U_{\mathbf{q}}=U_{\mathbf{q,}Ti}-U_{\mathbf{q,}O}\approx
\int_{uc}d^{3}re^{i\mathbf{qr}}(\frac{e^{\ast 2}Z_{Ti}^{\ast }}{|\mathbf{r-R}%
_{Ti}|}-\frac{e^{\ast 2}Z_{O}^{\ast }}{|\mathbf{r-R}_{O}|})$ . The effective
charges $Z_{\alpha }^{\ast }$ have been determined from a first principles
calculation \cite{Vanderbilt94} and are found to be unusually large in
perovskite compounds such as SrTiO$_{3}$, $Z_{Ti}^{\ast }=7.1$ and $%
Z_{O}^{\ast }=-5.7$. The reason is that the corresponding Ti-O ionic bond is
on the verge of \ being covalent, leading to large charge transfers. We thus
see that $U_{\mathbf{q}}$ comes out to be an order of magnitude larger than
usual deformation potentials. \ We account for the effect of electron
screening of the electron-ion potential by a factor $1/[1+(\kappa
/q_{typ})^{2}]$, with $\kappa $ the screening wave number and $%
q_{typ}\approx 1/a$ a typical wave number inside the first Brillouin zone ($%
a\approx 3.9\mathring{A}$\ is the lattice constant). This screening factor
causes a suppression of $V_{soft}$ with increasing electron density.

The momentum dependence of the matrix element of Bloch wave functions $u_{%
\mathbf{k}}(\mathbf{r)}$ (of a single band, for simplicity), integrated over
a unit cell of volume $V_{uc}$, $B_{\mathbf{q}}(\mathbf{k})=V_{uc}^{-1}\int
d^{3}ru_{\mathbf{k+q}}^{\ast }(\mathbf{r)}u_{\mathbf{k}}(\mathbf{r)}$ plays
an important role here, as it describes the symmetry components of the
coupling to the electron system. Separating out the average over the
Brillouin zone we define the anisotropic part $\Delta B_{\mathbf{q}}(\mathbf{%
k})$ of $B_{\mathbf{q}}(\mathbf{k})$ as $\Delta B_{\mathbf{q}}(\mathbf{k}%
)=B_{\mathbf{q}}(\mathbf{k})-\langle B_{\mathbf{q}}(\mathbf{k})\rangle _{av}$%
\ , where the average is over all $\mathbf{k}$ in the first Brillouin zone.
The coupling provided by the averaged $B$ is to density excitations of the
electron system, which are strongly screened, as seen above. As a
consequence, the frequncies of TO-phonon modes coupled to density
excitations are shifted up to the LO-phonon frequencies and are no longer
soft. In contrast, the anisotropic coupling mediated by $\Delta B_{\mathbf{q}%
}(\mathbf{k})$ does not couple the TO-phonon mode to density modes, leaving
the phonon frequency unchanged, i.e. soft. A microscopic calculation of $%
\Delta B_{\mathbf{q}}(\mathbf{k})$ requires an electronic band structure
calculation, which should take into account the substantial charge transfers
in the unit cell. The latter may be expected to give a strong anisotropic
component of $B_{\mathbf{q}}(\mathbf{k})$. We will assume $\Delta B_{\mathbf{%
q}}(\mathbf{k})$ of order unity in the following and approximate $(\Delta B_{%
\mathbf{q}}(\mathbf{k)})^{2}$\ by its average.

A crucial factor in Eq.(\ref{MTO}) is the inner product $(\mathbf{q\cdot e}%
_{TO}(\mathbf{q}))$ of the wave vector and the polarization direction of the
phonon mode. Along the principal directions $\mathbf{q}$ the polarization is
perpendicular to the propagation direction and the inner product vanishes.
In between these special directions $\mathbf{q}$ and $\mathbf{e}_{TO}(%
\mathbf{q})$ are not exactly orthogonal. The angular average of the e-ph
coupling function is therefore nonzero, if small. In Appendix B the angular
average $s=\langle (\widehat{\mathbf{q}}\mathbf{\cdot e}_{TO}(\mathbf{q}%
))^{2}\rangle $, where $\widehat{\mathbf{q}}=\mathbf{q}/|\mathbf{q}|$, is
calculated as $s\approx 0.1$, independent of $q$.\ Some evidence for a
coupling of electrons to the soft mode of SrTiO$_{3}$ has been found from
transport experiments \cite{Wemple66}.

The dimensionless TO phonon mediated interaction is then given by

\begin{equation}
N_{F}V_{soft}(\mathbf{q},i\omega _{n})=-\eta \frac{k_{F}}{q_{R}}\frac{q^{2}}{%
q_{R}^{2}}\frac{\omega _{D}^{2}}{\omega _{n}^{2}+\omega _{T}^{2}(q)}
\end{equation}%
where

\begin{equation}
\eta =\frac{2s}{\pi ^{2}}(q_{R}a)^{3}(\frac{\overline{U}}{\omega _{D}})^{2}%
\frac{m_{1}}{m_{ion}}\langle \Delta B_{\mathbf{q}}^{2}(\mathbf{k})\rangle
_{av}
\end{equation}%
Using the values $a=3.9\mathring{A}$\ , $q_{R}\approx 0.124\mathring{A}^{-1}$
, $\omega _{D}\approx 380$K , $\overline{U}\approx 35$eV, $%
m_{ion}/m_{1}=16320$ (taking the atomic mass of O$^{16}$),\ and $\langle
\Delta B^{2}\rangle \approx 1$,\ we estimate $\eta \approx 0.1$. The above
can provide only a rough estimate of the coupling strength $\eta $. A more
accurate determination requires a microscopic calculation, which is beyond
the scope of this work. We take a somewhat larger value, $\eta =0.18$ , in
the numerical evaluation, giving rise to the observed magnitude of the
isotope effect.

\section{Gap equation}

Given the pair potential, the equation determining the gap function $\Delta (%
\mathbf{k},i\omega _{n})$ in the spin singlet channel is given by \cite%
{Eliashberg}

\begin{eqnarray}
\Delta (\mathbf{k,}i\omega _{n}) &=&-T\sum_{\omega _{l}}\sum_{\mathbf{p\in }%
BZ}V_{pair}(\mathbf{k-p,}i\omega _{n}-i\omega _{l}) \\
&&\times \frac{\Delta (\mathbf{p,}i\omega _{l})}{[\omega _{l}-\Sigma (%
\mathbf{p,}i\omega _{l})]^{2}+\xi _{\mathbf{p}}^{2}+|\Delta (\mathbf{p,}%
i\omega _{l})|^{2}},  \notag
\end{eqnarray}%
where $\omega _{l}=(2l+1)\pi T$ is a fermionic Matsubara frequency, and $\xi
_{\mathbf{p}}=\epsilon _{\mathbf{p}}-\mu $ .\ Here $\Sigma (\mathbf{p}
,i\omega _{l})$\ is the normal self-energy, which is negligible at low $
\omega _{l}$, but will provide a cut-off at $|\omega _{l}|=\omega _{c}$ as
discussed below. The cutoff is determined by the condition $ Im \Sigma (
\mathbf{k}_{F},\omega _{c}-i0)=\omega _{c}$; at frequencies $|\omega
|>\omega _{c}$ quasiparticles are no longer well defined and the electron
spectral function drops rapidly to zero with growing $\omega $.

For small $V_{soft}, V_{pair}(\mathbf{q,}i\omega _{n})$ is a positive
definite function (on the real frequency axis $ Re V_{C}(\mathbf{q,}
\omega -i0)$ does have negative parts). The solutions $\Delta (\mathbf{k,}
i\omega)$ must therefore necessarily have negative components. This is
well known from the examples of spin fluctuation mediated $d$-wave
superconductivity, or of $p$-wave superfluidity of He$^{3}$, where the gap
parameter has nodes on the Fermi surface. We see here that it may also hold
for $s$-wave superconducting states. There the negative components must
arise in the frequency dependence. A negative part of  $\Delta $ at high
frequency was actually calculated by Morel and Anderson \cite{Morel-Anderson62}, 
and even before by Bogoliubov \cite{Bogoliubov58}. To
find such solutions requires to solve the gap equation as an integral
equation in frequency. For larger $V_{soft}$ (such as for $\eta =0.18 $, used
below)\ the pair interaction has sufficiently attractive (negative)
character that a solution without sign change in frequency becomes possible.
We restrict our considerations to the linear gap equation, which allows to
determine the transition temperature. It is expected that the e-ph
interaction gives rise to dominant s-wave pairing. Subdominant anisotropic
pairing, induced by the lattice anisotropy and the general momentum
dependence of the pair interaction, will not be considered here.

\subsection{Transition temperature}

Assuming a momentum independent gap function, the transition temperature $%
T_{c}$ follows from the linearized gap equation
\begin{equation}
\Delta (i\omega _{n})=-T_{c}\sum_{\omega _{l}}K(\omega _{n};\omega
_{l})\Delta (i\omega _{l}),
\end{equation}%
The kernel of the gap equation may be expressed as

\begin{eqnarray}
K(\omega _{n};\omega _{l}) &=&\int_{0}^{q_{c}}\frac{dqq^{2}}{2\pi ^{2}}%
V_{pair}(\mathbf{q},i\omega _{n}-i\omega _{l})F(q,i\omega _{l}),  \notag \\
F(q,i\omega _{l}) &=&\frac{1}{2}\int_{-1}^{1}\frac{d\cos \theta }{\omega
_{l}^{2}+(\xi _{\mathbf{k}_{F}}+\frac{q^{2}+2qk_{F}\cos \theta }{2m_{1}})^{2}%
},
\end{eqnarray}%
taking proper account of the cut-off induced by the self-energy. At small $%
\omega _{l}$ the function $F(q,i\omega _{l})\propto 1/\omega _{l}$, leading
to the wellknown logarithmic divergence of the kernel of the gap equation.
At large $\omega _{l}$ the self energy is dominant and $|\omega _{l}-\Sigma (%
\mathbf{p,}i\omega _{l})|\rightarrow \Sigma (\mathbf{p,}i\omega _{l})\propto
(\omega _{l})^{\beta }$, where $\beta =2$ or $3$, depending on density (see
Appendix A), thus providing the frequency cutoff at $\omega _{c}$. \ The
momentum integral may be done numerically or even analytically, employing
further approximations, with a cutoff $q_{c}$ of the order of $\pi /a$, the
extension of the first Brillouin zone. In view of the expected smallness of $%
T_{c}<<\omega _{phon}$ the frequency summation extends over thousands of
terms, allowing to replace the summation by an integral with a lower cutoff
at the smallest fermionic frequency, $\omega _{1}=\pi T$. It is useful to
consider even- frequency and odd-frequency solutions separately, by
introducing the corresponding kernels $K_{e,o}$. We define the
eigenfunctions $\psi _{\nu }^{e,o}(\omega )$ and eigenvalues $\alpha _{\nu
}^{e,o}(T_{c})$ of the kernels $K_{e,o}$ on the imaginary frequency axis as

\begin{eqnarray}
\alpha _{n}^{e,o}(T_{c})\psi _{n}^{e,o}(\omega ) &=&\int_{\pi T_{c}}^{\omega
_{c}}\frac{d\omega ^{\prime }}{2\pi }K_{e,o}(\omega ;\omega ^{\prime })\psi
_{n}^{e,o}(\omega ^{\prime }),  \notag \\
K_{e,o}(\omega ;\omega ^{\prime }) &=&K(\omega ;\omega ^{\prime })\pm
K(\omega ;-\omega ^{\prime })
\end{eqnarray}%
where the cutoff $\omega _{c}$ is found from an estimate of the imaginary
part of $\Sigma (\omega +i0)$ presented in the Appendix. We adopt the
interpolation expression \
\begin{equation}
\omega _{c}=\frac{\omega _{D}}{(c_{1}(\frac{k_{F}}{q_{R}})^{-1/2}+c_{2}(%
\frac{k_{F}}{q_{R}})^{2})^{-1}+c_{3}(\frac{k_{F}}{q_{R}})^{-1}}
\end{equation}%
with parameters $c_{1}$, $c_{2}$, $c_{3}$ (and where we defined $%
q_{R}^{2}/2m=\omega _{D}$ ). As shown in Appendix A one may distinguish
three density regimes with different dominant quasiparticle relaxation
processes: (i) a high density Fermi liquid regime with $\omega _{c}\approx
\epsilon _{F}$, (ii) an intermediate regime with dominant electron-electron
scattering and anomalous density dependence $\omega _{c}\approx \omega _{D}(%
\frac{k_{F}}{q_{R}})^{-1/2}$, and (iii) a low density regime where
electron-phonon scattering dominates, providing a cutoff $\omega _{c}\approx
30\omega _{D}(\frac{k_{F}}{q_{R}})$ $\gg \epsilon _{F}$. For typical
parameter values $\omega _{c}$ is a nonmonotonic function of density,
increasing at low density, passing through a maximum at around $n\approx
5\times 10^{17}$cm$^{-3}$ and through a minimum around $n\approx 2\times
10^{19}$cm$^{-3}$\ and increasing for higher densities.

The solution $\psi _{\nu }^{e,o}(\omega )$ with the highest transition
temperature may be obtained by finding the largest negative eigenvalue

\begin{equation}
\max_{\{n\}}[-\alpha _{n}^{e,o}(T_{c})]=-\alpha _{n_{0}}^{e,o}(T_{c})=1
\end{equation}%
and the gap function is given by $\Delta (i\omega )=\psi
_{n_{0}}^{e,o}(\omega )$.\ It is found that the highest transition
temperature appears in the even-frequency class for an eigenfunction $\psi
^{e}(\omega )$ with a single zero on the positive semi-axis (small $V_{soft}$%
) or without zero (sufficiently strong, but still small $V_{soft}$).

\begin{figure}[tbp]
\includegraphics[width=0.85\columnwidth]{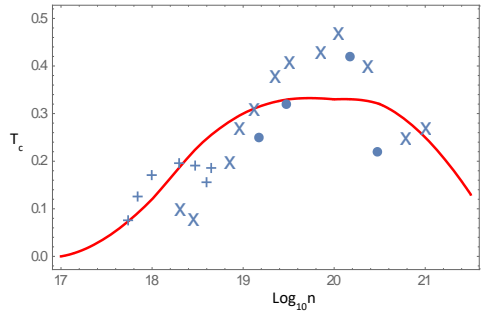}
\caption{Transition temperature Tc in Kelvin versus logarithm of electron
density in cm$^{-3}$. Theory: solid line; Experiment: crosses (Schooley et
al., 1964); filled circles: Nb-doped; + symbols: O-reduced (Lin et al.,
2014) }
\label{fig:Tc-logn}
\end{figure}

In Fig.\ \ref{fig:Tc-logn} we show the transition temperature $T_{c}$ as a
function of doping density $n$ . For a reasonable choice of the parameters
of the cutoff $\omega _{c}$, $c_{1}\approx 1.1$ , $c_{2}=0.6$ and $%
c_{3}\approx 0.036$, the $T_{c}$ values compare well with the experimental
data. For higher densities the higher electronic bands are successively
populated (gaps to the second band $4$meV and third band $30$meV, and
population of the second band starting at $n=3\times 10^{18}$cm$^{-3}$).
Their contribution is expected to increase $T_{c}$.

\subsection{Isotope effect}

\begin{figure}[tbp]
\includegraphics[width=0.85\columnwidth]{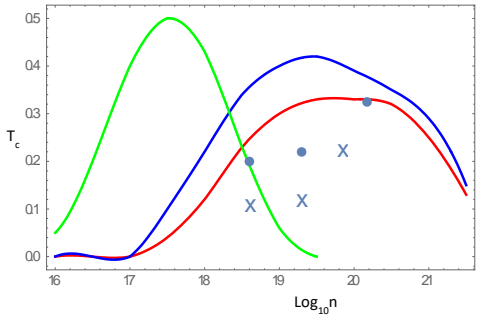}
\caption{Transition temperature Tc in Kelvin versus logarithm of electron
density in cm$^{-3}$. Theory: solid red line, isotope concentration x=0;
solid blue line, x=0.34; solid green line, only Coulomb interaction ($%
\protect\eta =0$) and x=0. Experiment: crosses, x=0; filled circles, x=0.34
(Stucky et al., 2016)). }
\label{fig:Tc-logn-iso}
\end{figure}

As already mentioned, a further spectacular finding about the
superconducting phase of doped SrTiO$_{3}$ is the observed isotope effect
\cite{vdMarel16}. One finds that substitution of O$^{16}$ by a concentration
$x_{c}=0.35$ of O$^{18}$ enhances $T_{c}$ by as much as $50\%$. In
conventional superconductors substitution by a heavier isotope leads to a
reduction of $T_{c}$ by a few percent, as caused by a slight decrease of the
prefactor in the BCS-expression for $T_{c}$. The isotope substitution moves
the system closer to the ferroelectric quantum critical point (or even
beyond it, into the ordered phase). This leads to a major change in the pair
interaction, through its dependence on the TO-phonon frequency $\omega
_{TO}(q=0)$, which vanishes at the QCP. Correspondingly, the pair attraction
is boosted by isotope substitution. We have calculated this effect by
putting $\omega _{TO}(q=0;x)=(1-x/x_{c})\omega _{TO}(q=0;x=0)$. As shown in
Fig.\ \ref{fig:Tc-logn-iso} we find that for the parameter specifying the
strength of $V_{soft}$, $\eta =0.18$ and keeping the cut-off frequency $%
\omega _{c}$ as presented above, $T_{c}$ is indeed increased by a factor $%
\approx 1.5$ at densities $n\approx 10^{18}$cm$^{-3}$ , with somewhat
smaller enhancement at higher densities. We point out that this approach is
different from the proposals by Edge et.al,\cite{Edge15,Kedem16}, where the
ferroelectric soft mode was discussed without taking into account the
Coulomb interactions. Also shown in Fig.\ \ref{fig:Tc-logn-iso} is the
result obtained without the TO-phonon mediated interaction ($\eta =0$).
Surprisingly, the $T_{c}$ curve is shifted to lower densities, but is not
lower, but even somewhat higher than the result obtained with both
interaction components ($\eta =0.18$).\ This clearly shows that the two
interaction components superpose in a more complex manner as one may have
thought, owing to their very different frequency dependence.

\section{Conclusion.}

Our results show that superconductivity in doped SrTiO$_{3}$ may be
interpreted as induced by two relevant interaction components, the
dynamically screened Coulomb interaction and an interaction mediated by
transverse optical phonons. Both, screening by ionic charges, in particular
through the optical phonon modes responsible for the incipient ferroelectric
transition and screening by electronic charges is important. At higher
densities electronic screening leads to a suppression of pairing. On the low
density side it is found that the rapid decay of quasiparticles for energies
beyond $\omega _{c}\propto n^{1/3}$ , caused by e-ph scattering confines the
domain of quasiparticle pairing to ever lower energies, as density
decreases, again leading to a suppression of $T_{c}$. An all-important
feature of our theory is the careful consideration of the relevant frequency
domain. We find that the cutoff frequency at low density is much larger than
the Fermi energy, and even varies with density in a nonmonotonic fashion.
Previous studies on, e.g. plasmon exchange mediated superconductivity\cite%
{Rietschel83} have found a strong effect of higher order contributions,
substantially reducing the tendency for superconductivity\cite{Rietschel90}.
In the present case higher order corrections are small because both, the
screened Coulomb interaction and the soft phonon induced interaction are
weak. The observation of a large and unusual isotope effect may be explained
easily within the present model: isotope substitution can move the system
closer to the ferroelectric transition, and thereby softens the TO-phonon
frequency. This leads to a substantial enhancement of the pairing
interaction.

The inclusion of higher phonon modes is not expected to lead to major
changes, as it will only affect the high frequency part of the pairing
interaction. In contrast, taking into account the two higher electronic
bands, which begin to be populated at higher dopings is expected to increase
the pairing tendency sizeably.

\section{Acknowledgments}

We acknowledge useful discussions with Kirsty Dunnett, Yaron Kedem, Thilo
Kopp, Jeremy Levy, J\"{o}rg Schmalian, Nicola Spaldin and Ilya Sochnikov. We
thank Patrick Lee and Jonathan Ruhman for constructive criticism. PW thanks
Boris Shklovskii and Konstantin Reich for sharing their knowledge of the
screening properties of SrTiO$_{3}$ (see Appendix B) and for discussions on
possible d-wave superconductivity. PW is also grateful to the theory group
at the Institute for Materials Science, Los Alamos National Laboratory, for
hospitality and especially for assistance with the numerical solution of the
gap equation in the initial stages by Zhoushen Huang and Towfiq Ahmed. Work
is supported by VILLUM FONDEN via the Centre of Excellence for Dirac
Materials (Grant No. 11744), Knut and Alice Wallenberg Foundation and the
European Research Council under the European Union Seventh Framework Program
(FP/2207-2013)/ERC Grant Agreement No. DM- 32103. Part of this work has been
performed at the Aspen Center for Physics, which is supported by NSF Grant
No. PHY-1066293.

\section{Appendices}

\subsection{Electronic self-energy}

The purpose of this Appendix is to show that the quasiparticle relaxation
rate $\Gamma $ exceeds the energy $\omega $ for $|\omega |>\omega _{c}$ ,
signaling that quasiparticles are no longer well defined and that the
electron spectral function falls off faster than $\omega ^{-2}$.
Consequently the contribution of processes from the energy domain $|\omega
|>\omega _{c}$ to the gap equation is strongly suppressed, even though the
screened Coulomb interaction is perfectly finite for such energies.

\subsubsection{Coulomb interaction}

We first estimate the contribution to $\Gamma (\mathbf{k}_{F},\omega )=
Im \Sigma (\mathbf{k}_{F},\omega -i0)$ from electron-electron interaction, $
\Gamma _{el}$, as approximately given by
\begin{eqnarray}
\Gamma _{el}(\mathbf{k}_{F},\omega ) &=&\pi \sum_{\mathbf{k}_{2},\mathbf{k}%
_{3}}\left\vert V(\left\vert \mathbf{k}_{F}-\mathbf{k}_{3}\right\vert
,\omega -\epsilon _{\mathbf{k}_{3}})\right\vert ^{2}  \notag \\
&&\times (1-n_{\mathbf{k}_{3}})(1-n_{\mathbf{k}_{4}})n_{\mathbf{k}%
_{2}}\delta _{\epsilon }
\end{eqnarray}%
where $\mathbf{k}_{4}=\mathbf{k}_{F}+\mathbf{k}_{2}-\mathbf{k}_{3}$ and $%
\delta _{\epsilon }=\delta (\omega +\epsilon _{\mathbf{k}_{2}}-\epsilon _{%
\mathbf{k}_{3}}-\epsilon _{\mathbf{k}_{4}})$.

Let us first look at energy $\omega \approx \epsilon _{F}$, at the border of
the Fermi liquid regime, when we have

\begin{equation}
\Gamma _{el}(\mathbf{k}_{F},\epsilon _{F})\approx \frac{\omega ^{2}}{%
\epsilon _{F}}\left\vert N_{0}V_{C}(\left\vert \mathbf{k}_{F}\right\vert
,\epsilon _{F})\right\vert ^{2}
\end{equation}%
In the limit of high densities, where $\epsilon _{F}\gtrsim \omega
_{LO}(q=0) $ we have

\begin{equation}
N_{0}V_{C}(\left\vert \mathbf{k}_{F}\right\vert ,\epsilon _{F})\approx \frac{%
\kappa ^{2}}{k_{F}^{2}+\kappa ^{2}}\approx 1
\end{equation}%
so that the cutoff frequency is given by $\omega _{c1}\approx \epsilon _{F}$%
. \

At lower densities $\epsilon _{F}\ll \omega _{LO}(q=0)$ the screening
provided by the ionic charges strongly weakens the Coulomb interaction. We
have approximately

\begin{eqnarray}
N_{0}V_{C}(\left\vert \mathbf{k}_{F}\right\vert ,\epsilon _{F}) &\approx &%
\frac{\kappa ^{2}}{k_{F}^{2}\frac{\omega _{LO,0}^{2}(q=0)}{\epsilon
_{F}^{2}+\omega _{TO}^{2}(q=0)}+\kappa ^{2}}  \notag \\
&\approx &\frac{\kappa ^{2}}{k_{F}^{2}}\frac{\epsilon _{F}^{2}+\omega
_{TO}^{2}(q=0)}{\omega _{LO,0}^{2}(q=0)}
\end{eqnarray}%
At a density of $n=10^{18}$cm$^{-3}$ , putting in numbers $k_{F}\approx 0.03%
\mathring{A}^{-1}$, $\kappa ^{2}\approx 0.012\mathring{A}^{-2}$ and $\omega
_{TO}^{2}(q=0)/\omega _{LO,0}^{2}(q=0)\approx 10^{-3}$ it is apparent that $%
\gamma =\left\vert N_{0}V_{C}(\left\vert \mathbf{k}_{F}\right\vert ,\epsilon
_{F})\right\vert ^{2}\approx 10^{-4}\ll 1$. Therefore, the resulting cutoff
frequency will be $\omega _{c}\gg \epsilon _{F}$.

For large $\omega \gg \epsilon _{F}$\ energy conservation requires the qp
energies $\epsilon _{3,4}$ to be large such that $n_{\mathbf{k}_{3}},n_{%
\mathbf{k}_{4}}\ll 1$ . The momentum integrations may be expressed in terms
of integrations over the energies $\epsilon _{\mathbf{k}_{2}},\epsilon _{%
\mathbf{k}_{3}}$ and over angles, in particular over $\cos \theta =(\mathbf{k%
}_{F}\cdot \mathbf{k}_{3})/k_{F}k_{3}$

\begin{eqnarray}
\Gamma _{el}(\mathbf{k}_{F},\omega ) &\approx &\pi \int_{0}^{k_{F}}\frac{%
dk_{2}k_{2}^{2}}{2\pi ^{2}}\int \frac{d\Omega _{\mathbf{k}_{2}}}{4\pi }%
\int_{0}^{\pi /a}\frac{dk_{3}k_{3}^{2}}{2\pi ^{2}}  \notag \\
&&\times \int \frac{d\cos \theta d\phi }{4\pi }\left\vert V_{C}(\left\vert
\mathbf{k}_{F}-\mathbf{k}_{3}\right\vert ,\omega -\epsilon _{3})\right\vert
^{2}\delta _{\Gamma }.
\end{eqnarray}%
Here $\delta _{\Gamma }=\delta _{\Gamma }(\omega +\epsilon _{2}-\epsilon
_{3}-\epsilon _{4})$ is a "delta function" of finite width $\Gamma $,
allowing for the fact that $\Gamma $ may be as large as the energy itself. \
The integral over $\cos \theta $ may be done with the help of the $\delta -$%
function, using $\epsilon _{4}\approx v|\mathbf{k}_{4}|=v(\Delta \epsilon
-2k_{F}k_{3}\cos \theta )^{1/2}$ , for $\mathbf{k}_{4}$ inside the first
Brillouin zone. We expect $\epsilon _{4}$ to be large, beyond the regime
where the quadratic dependence on $\mathbf{k}_{4}$ holds. We therefore use a
rough approximation of the electronic energy dispersion for higher energies,
replacing so to speak the $-\cos k$ -function of a tight-binding dispersion
by a straight line. The result is
\begin{eqnarray}
\Gamma _{el}(\mathbf{k}_{F},\omega ) &\propto &\int_{0}^{k_{F}}\frac{%
dk_{2}k_{2}^{2}}{2\pi ^{2}}\int_{0}^{\infty }\frac{dk_{3}k_{3}^{2}}{2\pi ^{2}%
}  \notag \\
&&\times \frac{\omega }{k_{F}k_{3}v^{2}}\left\vert V_{C}(\left\vert \mathbf{k%
}_{3}\right\vert ,\omega -\epsilon _{\mathbf{k}_{3}})\right\vert ^{2},
\end{eqnarray}%
where we used $\epsilon _{\mathbf{k}_{2}}\ll \epsilon _{\mathbf{k}_{3}}$ and
averaged over the remaining angles. What is the screened Coulomb interaction
in the range $\omega _{TO}(q_{typ}),\epsilon _{F}\ll \omega \ll \omega
_{LO}(q_{typ},\omega )$ and for typical values of momentum $q_{typ}\approx
\omega /v$? Because of the large difference of the transverse and the
longitudinal optical phonon frequencies in a polar material, there is a
frequency regime where $V_{C}\propto \omega ^{2}$ , reflecting the huge
difference in polarization at low and high frequencies. We may approximate
the Coulomb potential by

\begin{eqnarray}
V_{C}(q,\omega ) &=&\frac{4\pi e^{\ast 2}}{q^{2}\frac{\omega ^{2}-\omega
_{LO,0}^{2}(k_{3})}{\omega ^{2}-\omega _{TO}^{2}(k_{3})}+\kappa ^{2}}  \notag
\\
&\approx &\frac{4\pi e^{\ast 2}}{k_{3}^{2}+\kappa _{\omega }^{2}}\frac{%
\omega ^{2}}{\omega _{D}^{2}}
\end{eqnarray}%
where $\kappa _{\omega }^{2}=\kappa ^{2}\omega ^{2}/\omega _{D}^{2}$\ .
Within the above approximations one finds

\begin{eqnarray}
\Gamma _{el}(\mathbf{k}_{F},\omega ) &\propto &\frac{4k_{F}^{2}e^{\ast 4}}{%
\pi ^{2}v^{2}}\frac{\omega ^{5}}{\omega _{D}^{4}}\int_{\kappa _{\omega
}}^{\infty }dk_{3}\frac{1}{k_{3}^{3}}\approx \omega \left( \frac{\omega }{%
\omega _{c2}}\right) ^{2},  \notag \\
\omega _{c2} &\approx &\pi \omega _{D}\frac{(mv^{2})^{1/2}}{(e^{\ast
2}q_{R})^{1/2}}(q_{R}/k_{F})^{1/2}
\end{eqnarray}%
Using that $v\approx D/(\pi /a)$ , $(\pi /a)^{2}/m\approx D$\ and $e^{\ast
2}/a\approx D$, where $D$ is the half band width, it is seen that the ratio $%
(mv^{2})^{1/2}/(e^{\ast 2}/a)^{1/2}=O(1)$. The cutoff frequency is seen to
decrease weakly with increasing $k_{F}$, in the intermediate density regime.

\subsubsection{Electron-phonon scattering}

At small densities the contribution from phonon scattering to $\Gamma $
dominates. We consider scattering by acoustical phonons, approximating the
dispersion by $\omega _{\mathbf{q}}=cq$ and taking into account that the
polarization vector $\mathbf{e}_{\lambda }$ is parallel to $\mathbf{q}$
along the principal directions. The scattering rate may be estimated as
\begin{equation}
\Gamma _{ph}(\mathbf{k}_{F},\omega )\approx \pi \sum_{\mathbf{q}}\left\vert
M(\mathbf{q})\right\vert ^{2}(1-n_{\mathbf{k+q}})\delta _{\Gamma }(\omega
-\epsilon _{\mathbf{k+q}}-\omega _{\mathbf{q}}),
\end{equation}%
We approximate the electronic dispersion by $\epsilon _{\mathbf{k}}\approx
v(|\mathbf{k|-}k_{F})$ .\ Using $n_{\mathbf{k+q}}=\Theta (\omega _{\mathbf{q}%
}-\omega )$\ the angular integral yields

\begin{eqnarray}
\Gamma _{ph}(\mathbf{k}_{F},\omega ) &\approx &\int \frac{dqq^{2}}{2\pi ^{2}}%
\langle \left\vert M(\mathbf{q})\right\vert ^{2}\rangle \Theta (\omega
-\omega _{q})\frac{\omega -\omega _{q}}{k_{F}qv^{2}}  \notag \\
&&\times \Theta _{\Gamma }(2k_{F}q-|(\frac{\omega }{v})^{2}-q^{2}|).
\end{eqnarray}%
where $\Theta _{\Gamma }(x)$ is a step function of width $\Gamma $.\ Here
the e-ph matrix element squared is approximated by

\begin{equation}
\langle \left\vert M(\mathbf{q})\right\vert ^{2}\rangle \approx \frac{q^{2}}{%
\rho \omega _{q}}(\frac{4\pi e^{\ast 2}}{a^{3}(q^{2}+\kappa ^{2})})^{2},
\end{equation}%
where $\rho =m_{ion}a^{-3}$ is the ionic mass density and $a$ is the lattice
spacing. Here we are allowed to drop the ionic part of the screening,
considering the fact that the relevant momentum transfer $q_{typ}\approx
\omega _{D}/v$\ and hence $\omega _{TO}(q_{typ})\approx \omega
_{LO}(q_{typ},\omega _{D})$ .\ We have $vq\gg \omega _{\mathbf{q}}$, so that
we may drop $\omega _{\mathbf{q}}$ from the argument of the second
step-function, also $k_{F}^{2}\ll q^{2}$\ may be neglected. The $q-$integral
is then confined to a region of width $2k_{F}$ around $q_{\omega }=\omega /v$%
. As a result we get at $\omega <v\kappa $

\begin{equation}
\Gamma _{ph}(\mathbf{k}_{F},\omega )\approx \omega ^{3}\frac{1}{\pi
^{2}m_{ion}ca}(\frac{4\pi e^{\ast 2}}{av^{2}\kappa ^{2}})^{2}=\omega (\frac{%
\omega }{\omega _{c3}})^{2},
\end{equation}%
Estimating $v\approx \pi /(m_{1}a)$, $c^{2}/v^{2}\approx m_{1}/m_{ion}$, and
using $4\pi e^{\ast 2}/\kappa ^{2}=\pi ^{2}/m_{1}k_{F}$ one finds

\begin{equation}
\omega _{c3}\approx \omega _{D}(\frac{mv^{2}}{\pi ^{2}\omega _{D}})^{1/2}(%
\frac{m_{ion}}{m_{1}})^{1/4}(\frac{k_{F}}{q_{R}})
\end{equation}%
\ The two prefactors may be estimated as $(\frac{mv^{2}}{\omega _{D}}%
)^{1/2}\approx 5$ and $(\frac{m_{ion}}{m_{1}})^{1/4}\approx 13$ (using the
ionic mass of O$^{16}$), yielding an estimate of the parameter $c_{3}$
introduced in Eq.(14), $c_{3}\approx 0.05$ , which compares well with the
value used in the numerical evaluation.

\subsubsection{Cutoff energy}

Combining the above results for the cutoff energies induced by Coulomb
interaction at high and low densities, $\omega _{c1},\omega _{c2}$ and by
the electron-phonon interaction, $\omega _{c3}$ into a single expression we
may define

\begin{equation}
\omega _{c}=\frac{\omega _{D}}{(c_{1}(\frac{k_{F}}{q_{R}})^{-1/2}+c_{2}(%
\frac{k_{F}}{q_{R}})^{2})^{-1}+c_{3}(\frac{k_{F}}{q_{R}})^{-1}}
\end{equation}%
For the values of the parameters $c_{1},c_{2},c_{3}$ used in the numerical
evaluation, the cutoff energy is found to be a nonmonotonic function of $%
k_{F}$, which may be traced to $\omega _{c2}\propto k_{F}^{-1/2}$,
decreasing with increasing density, mainly because $\Gamma \propto n/\kappa
^{4}\propto k_{F}$ . This nonmonotonic behavior appears to be necessary for
obtaining the observed $T_{c\text{ }}$versus density values.

\subsection{Phenomenological model of optical phonons of SrTiO$_{3}$}

\subsubsection{Optical phonons in a dynamical model of electric polarization}

The soft mode properties of SrTiO$_{3}$ may be discussed in the framework of
a Ginzburg-Landau-Wilson action for the electric polarization $\mathbf{P(r}%
,t)$ , varying in space and imaginary time. The general form of the action
(dropping nonlinear terms) is

\begin{equation}
S=\frac{1}{2}\int dt\int d\mathbf{r[}\frac{1}{\omega _{D}^{2}}(\frac{%
\partial \mathbf{P}}{\partial t})^{2}+\tau \mathbf{P}^{2}\mathbf{+}%
\sum_{i,j,k,l}\lambda _{ijkl}\frac{\partial P_{i}}{\partial x_{j}}\frac{%
\partial P_{k}}{\partial x_{l}}].
\end{equation}%
The fourth rank tensor $\lambda _{ijkl}$ in a cubic lattice (we neglect the
tetragonal distortion for simplicity) has three independent elements $%
\lambda _{j}$, $j=1,2,3$

\begin{equation}
\lambda _{ijkl}=\lambda _{1}\delta _{ik}\delta _{kl}\delta _{ij}+\lambda
_{2}\delta _{kl}\delta _{ij}(1-\delta _{ik})+\lambda _{3}\delta _{ik}\delta
_{jl}(1-\delta _{ij}).
\end{equation}%
Stability requires $\lambda _{1}>\lambda _{2}>0$ and $\lambda _{3}>0$.\ The
electric field $\mathbf{E}$\ generated by $\mathbf{P}$ is obtained as

\begin{equation}
\mathbf{E(r},t)=\frac{\delta S}{\delta \mathbf{P(r},t)}=\mathbf{\chi }\cdot
\mathbf{P},
\end{equation}%
where $\mathbf{\chi }$ is the tensor of electric susceptibility, the
elements of which are defined in Fourier space as

\begin{equation}
\chi _{ik}(\mathbf{q},i\omega _{n})=(\frac{\omega _{n}^{2}}{\omega _{D}^{2}}%
+\tau )\delta _{ik}+q^{2}A_{ik}
\end{equation}%
where we defined a tensor

\begin{equation}
A_{ik}=[\lambda _{3}(1-n_{i}^{2})+\lambda _{1}n_{i}^{2}]\delta _{ik}+\lambda
_{2}n_{i}n_{k}(1-\delta _{ik})
\end{equation}%
and $\widehat{\mathbf{q}}=(n_{1},n_{2},n_{3})$ is the unit vector in the
direction of $\mathbf{q}$.

To determine the Fourier components of the potential $\varphi _{ion}(\mathbf{%
r},t)$\ of a point charge screened by ionic charges we need to solve the
equations

\begin{align}
& i\mathbf{q\cdot (E+}4\pi \mathbf{P})=4\pi e  \notag \\
\mathbf{E}& \mathbf{=}-i\mathbf{q}\varphi _{ion}=\mathbf{\chi \cdot P}
\notag
\end{align}%
with the result

\begin{equation}
\varphi _{ion}(\mathbf{q,}i\omega _{n})=\frac{4\pi e}{q^{2}+4\pi \mathbf{%
q\cdot \chi }^{-1}\mathbf{\cdot q}}=\frac{4\pi e}{\epsilon _{ion}(\mathbf{q}%
,i\omega _{n})q^{2}}
\end{equation}%
The projection of the inverse susceptibility along $\mathbf{q}$ is
approximately given by

\begin{equation}
\mathbf{q\cdot \chi }^{-1}\mathbf{\cdot q}\approx \frac{q^{2}}{\frac{\omega
_{n}^{2}}{\omega _{D}^{2}}+\tau +\lambda _{1}q^{2}+2K(-\lambda _{1}+\lambda
_{3}+\lambda _{2})q^{2}}  \label{chi}
\end{equation}%
where $K(\widehat{\mathbf{q}}%
)=n_{x}^{2}n_{y}^{2}+n_{y}^{2}n_{z}^{2}+n_{z}^{2}n_{x}^{2}$. Eq.(\ref{chi})
is exact for $\mathbf{q}$ along the axes, face diagonals or space diagonals.
It follows that

\begin{equation}
\epsilon _{ion}\mathbf{(q},i\omega _{n})\approx \epsilon _{\infty }\frac{%
\omega _{n}^{2}+\omega _{LO,0}^{2}(\mathbf{q})}{\omega _{n}^{2}+\omega
_{TO}^{2}(\mathbf{q})}
\end{equation}%
where we defined the average frequencies of the soft TO phonon and the
accompanying (bare) LO phonon by

\begin{eqnarray}
\frac{\omega _{TO}^{2}(\mathbf{q})}{\omega _{D}^{2}} &=&\tau +\lambda
_{1}q^{2}+2K(-\lambda _{1}+\lambda _{3}+\lambda _{2})q^{2}, \\
\omega _{LO,0}^{2}(\mathbf{q}) &\approx &\omega _{TO}^{2}(\mathbf{q})+4\pi
\omega _{D}^{2}
\end{eqnarray}%
The exact TO phonon frequencies are obtained from the zeros of the
eigenvalues of the tensor $\mathbf{\chi }$ (see below).

\subsubsection{Dielectric function of doped SrTiO$_{3}$\ }

At finite doping the potential $\varphi $ of\ a test charge is additionally
screened by the conduction electron system

\begin{equation}
e\varphi (\mathbf{q},i\omega _{n})=\frac{e\varphi _{ion}}{1+e\varphi
_{ion}\chi _{el}}=\frac{4\pi e^{2}}{\epsilon (\mathbf{q},i\omega _{n})q^{2}}
\end{equation}%
Here the dielectric function is given by

\begin{equation}
\epsilon (\mathbf{q},i\omega _{n})=\epsilon _{\infty }[1+\frac{4\pi \mathbf{%
q\cdot \chi (q},\omega _{n})^{-1}\mathbf{\cdot q}}{q^{2}}+\frac{4\pi e^{2}}{%
\epsilon _{\infty }q^{2}}\chi _{el}\mathbf{(q},i\omega _{n})]
\end{equation}%
where $\chi _{el}$ is the irreducible electric polarization of the
conduction electrons. The potential of a fully screened test charge is
equivalent to the screened Coulomb interaction $V_{C}(\mathbf{q},i\omega
_{n})$ between two conduction electrons, which may be reexpressed as

\begin{equation}
V_{C}(\mathbf{q},i\omega _{n})=\frac{4\pi e^{\ast 2}}{q^{2}+\kappa ^{2}(%
\mathbf{q},i\omega _{n})}\frac{\omega _{n}^{2}+\omega _{TO}^{2}(\mathbf{q})}{%
\omega _{n}^{2}+\omega _{LO}^{2}(\mathbf{q},i\omega _{n})},
\end{equation}%
where $\kappa ^{2}(\mathbf{q},i\omega _{n})=4\pi e^{\ast 2}\chi _{el}(%
\mathbf{q},i\omega _{n})$ and $\omega _{LO}(\mathbf{q},i\omega _{n})$ is the
LO phonon frequency renormalized by electronic screening effects

\begin{equation}
\omega _{LO}^{2}(\mathbf{q},i\omega _{n})=\omega _{TO}^{2}(\mathbf{q})+\frac{%
4\pi q^{2}}{q^{2}+\kappa ^{2}(\mathbf{q},i\omega _{n})}\omega _{D}^{2}.
\end{equation}%
To make contact with the usual representation of the phonon mediated
interaction we note that $V_{C}$ may be expressed as the sum of an
electronically screened Coulomb interaction and a phonon induced interaction

\begin{eqnarray}
V_{C}(\mathbf{q},i\omega _{n}) &=&\frac{4\pi e^{\ast 2}}{q^{2}+\kappa ^{2}(%
\mathbf{q},i\omega _{n})} \\
&&-\frac{4\pi e^{\ast 2}q^{2}}{(q^{2}+\kappa ^{2}(\mathbf{q},i\omega
_{n}))^{2}}\frac{4\pi \omega _{D}^{2}}{\omega _{n}^{2}+\omega _{LO}^{2}(%
\mathbf{q},i\omega _{n})}.  \notag
\end{eqnarray}

\subsubsection{Transverse optical phonon eigenstates}

We define the eigenstates of the tensor $A_{jk}$ by

\begin{equation}
\sum_{k}A_{jk}e_{k}^{(m)}=a^{(m)}e_{j}^{(m)}.
\end{equation}%
The transverse optical phonon frequencies and polarization vectors are found
as solutions of $\mathbf{E(q},\omega _{n})=\mathbf{\chi (q},\omega
_{n})\cdot \mathbf{P(q},\omega _{n})=0$, at finite $\mathbf{P}$ and for
transverse polarization. Consequently we are looking for eigenstates of the
tensor $\mathbf{\chi }$ with zero eigenvalue, such that

\begin{equation}
\frac{(i\omega _{n})^{2}}{\omega _{D}^{2}}=\tau +q^{2}a^{(m)}
\end{equation}%
and eigenvectors $e_{j}^{(m)}$. We have calculated the eigenvalues and
eigenstates of $\mathbf{\chi }$ in their dependence on the direction $%
\widehat{\mathbf{q}}$\ , and hence the TO soft phonon frequency and
eigenvector (identified as the lowest eigenvalue). For orientations $%
\widehat{\mathbf{q}}=(1,0,0)$ and $\widehat{\mathbf{e}}^{(1)}=(0,1,0)$ we
find $a^{(1)}=\lambda _{3}$ , which compared with INS data \cite{Yamada69}\
yields a value $\lambda _{3}\approx 2.0\mathring{A}^{2}$. For the only other
experimental configuration of $\widehat{\mathbf{q}}=\frac{1}{\sqrt{2}}%
(1,1,0) $ and $\widehat{\mathbf{e}}^{(1)}=\frac{1}{\sqrt{2}}(1,-1,0)$ we get
$a^{(1)}=\frac{1}{2}(\lambda _{1}-\lambda _{2}+\lambda _{3})$\ , compared
with the experimental value \cite{Yamada69} $a^{(1)}\approx 4.5\mathring{A}%
^{2}$. Assuming a value of $\lambda _{2}\approx 1.0\mathring{A}^{2}$ one
then obtains $\lambda _{1}\approx 8.0\mathring{A}^{2}$ .

Using these values we have determined the eigenstates of the matrix $A_{jk}$
numerically.\ The angular average of the scalar product of the unit vector
along $\mathbf{q}$ and an eigenvector $\mathbf{e}^{(TO)}$ squared is found as

\begin{equation}
\langle (\widehat{\mathbf{q}}\cdot \mathbf{e}^{(TO)})^{2}\rangle \approx 0.1
\end{equation}

\subsection{Vertex corrections}

Here we provide a rough estimate of the vertex correction in first order of
the interaction. It is given by

\begin{equation}
\Lambda (k,p)=\sum_{q}G(k-q)G(p-q)V(q)
\end{equation}%
where $G(k)=G(\mathbf{k},i\omega _{n})=(i\omega _{n}-\xi _{\mathbf{k}})^{-1}$%
, omitting the self-energy correction, and $\xi _{\mathbf{k}}=\epsilon _{%
\mathbf{k}}-\mu $. We first consider the screened Coulomb interaction, using
the approximation

\begin{equation}
V_{C}(\mathbf{q},i\omega _{n})\approx \frac{4\pi e^{\ast 2}}{q^{2}+\kappa
_{0}^{2}}\frac{\omega _{TO}^{2}(\mathbf{\kappa }_{0})}{2\pi \omega _{D}^{2}}.
\end{equation}%
Here $\kappa _{0}^{2}=\kappa ^{2}(\mathbf{0},0)=4\pi e^{\ast 2}N_{F}$, which
at a density of $n=10^{18}$cm$^{-3}$ amounts to $\kappa _{0}^{2}\approx 0.026%
\mathring{A}^{-2}$. It follows that the frequency ratio is small, $\rho
=\omega _{TO}^{2}(\mathbf{\kappa }_{0})/2\pi \omega _{D}^{2}=(\tau +\lambda
\kappa _{0}^{2})/2\pi \approx 0.018$, so that the dimensionless interaction
strength $N_{F}V_{C}\approx 10^{-2}$ in the relevant regime of momenta and
frequencies. The vertex corrections are approximately given by (using $\xi _{%
\mathbf{q}}=v_{F}(q-k_{F})$, where $v_{F}=k_{F}/m_{1}$ is the Fermi
velocity, and neglecting $k,p$ compared to $q$ )
\begin{equation}
\Lambda _{C}(k,p)\approx N_{F}^{-1}\rho T\sum_{\omega _{l}}\int \frac{dqq^{2}%
}{2\pi ^{2}}\frac{\kappa _{0}^{2}}{q^{2}+\kappa _{0}^{2}}\frac{1}{(i\omega
_{l}-\xi _{\mathbf{q}})^{2}}\approx \rho
\end{equation}%
where the frequency summation has been done approximately as

\begin{eqnarray}
T\sum_{\omega _{l}}\frac{1}{(i\omega _{l}-\epsilon _{q})^{2}} &\approx
&\{\int_{\pi T}^{\infty }+\int_{-\infty }^{-\pi T}\}\frac{d\omega }{2\pi }%
\frac{1}{(i\omega -\xi _{\mathbf{q}})^{2}} \\
&\approx &\frac{T}{(\pi T)^{2}+\xi _{\mathbf{q}}^{2}{}}  \notag
\end{eqnarray}%
and we used $T\ll \epsilon _{F}$ and $\kappa _{0}^{2}+q^{2}\approx \kappa
_{0}^{2}$, since $q\approx k_{F}\ll \kappa _{0}$\ .

The TO-phonon mediated interaction gives rise to the following vertex
correction

\begin{eqnarray}
\Lambda _{soft}(k,p) &\approx &\frac{\eta }{N_{F}}\frac{k_{F}}{q_{R}}%
T\sum_{\omega _{l}}\int \frac{dqq^{2}}{2\pi ^{2}}\frac{q^{2}}{q_{R}^{2}}%
\frac{\omega _{D}^{2}}{\omega _{T}^{2}(q)}\frac{1}{(i\omega _{l}-\xi _{%
\mathbf{q}})^{2}}  \notag \\
&\approx &\frac{\eta }{N_{F}}\frac{k_{F}}{q_{R}}\int \frac{dqq^{2}}{2\pi ^{2}%
}\frac{q^{2}}{q_{R}^{2}}\frac{\omega _{D}^{2}}{\omega _{T}^{2}(q)}\frac{T}{%
(\pi T)^{2}+\xi _{\mathbf{q}}^{2}}
\end{eqnarray}%
where the frequency summation has been done as described above. The momentum
integral is again dominated by the sharp peak at $\epsilon _{q}=0$, i.e. at $%
q=k_{F},$ and may be approximately done as

\begin{equation}
\Lambda _{soft}(k,p)\approx \eta (\frac{k_{F}}{q_{R}})^{3}\frac{1}{\tau
+\lambda k_{F}^{2}}
\end{equation}
At a density of \ $n=10^{18}$cm$^{-3}$ we may estimate $k_{F}\approx 0.03%
\mathring{A}^{-1}$, therefore $\eta (k_{F}/q_{R})^{3}\approx 2\times 10^{-3}$
(we recall $q_{R}\approx 0.124\mathring{A}^{-1}$). Taking $(\tau +\lambda
k_{F}^{2})^{-1}\approx 100$ we then find $\Lambda _{soft}\approx 0.2$.

These rough estimates demonstrates that vertex corrections are negligible on
account of the unusually strong screening by ionic charges in this
electrically highly polarizable material and due to the weak coupling of
electrons to TO phonons.

\end{document}